\documentclass[a4paper,11pt]{article}
\usepackage{jheppub} 
\usepackage{lineno}
\usepackage{txfonts}
\usepackage{physics}
\usepackage{subcaption}
\usepackage{mathrsfs}
\usepackage{amsmath}
\usepackage{tikz}
\usetikzlibrary{arrows,matrix,positioning}
\usepackage{appendix}

\title{Note on the local calculation of decoherence of quantum superpositions in de Sitter spacetime}

\author{Ran Li}

\affiliation{Department of Physics, Qufu Normal University, Qufu, Shandong 273165, China}

\emailAdd{liran@qfnu.edu.cn}

\abstract{We study the decoherence effect of quantum superposition in de Sitter (dS) spacetime due to the presence of the cosmological horizon. Using the algebraic approach of quantum field theory on curved spacetime, we derive the precise expression for the expected number of entangling particles in the scalar field case. This expression establishes the relation between the decoherence and the local two-point correlation function. Specifically, we analyze the quantum superposition Gendankenexperiment performed by a local observer at the center of dS spacetime. We compute the entangling particle numbers in scalar field, electromagnetic field, and gravitational field scenarios. It is demonstrated that the quantum spatial superposition state can be decohered by emitting entangling particles into the cosmological horizon. Our setup is equivalent to an accelerating observer in 5-dimensional Minkowski spacetime. The results for the scalar and electromagnetic cases are consistent with those obtained in \cite{Wilson-Gerow:2024ljx}, which investigated the decoherence effect from the perspective of an accelerating observer in Minkovski spacetime. However, our result fixes the numerical prefactor of the gravitational decoherence. }

\begin{document}

\maketitle

\section{Introduction}
\label{sec:intro}

The unification of quantum mechanics and gravitation is one of the most challenging and ambitious open problems in modern fundamental physics. String theory and loop quantum gravity are two prominent, mathematically sophisticated frameworks that aim to quantize gravity. However, an alternative approach takes the opposite perspective: to ``gravitize" quantum mechanics \cite{Penrose:2014nha}. In this approach, it is assumed that the quantum system interacts with classical or quantum gravitational backgrounds, offering a different route to understanding the interplay between these foundational theories \cite{Bassi:2017szd}. Although a complete theory of quantum gravity has yet to be established, there is a general consensus that spacetime undergoes fluctuations \cite{Hu:2008rga}. In addition, unlike the electromagnetic interaction, the gravitational interaction cannot be shielded. As a result, spacetime, as a dynamical variable, fluctuates, and these fluctuations must lead to decoherence in quantum systems \cite{Anastopoulos:2021jdz}.

Decoherence is a fundamental concept in quantum mechanics that describes how a quantum system initially in a superposition state loses its coherence due to interactions with the environment \cite{Zurek:2003zz,Schlosshauer:2003zy,Schlosshauer:2019ewh}. It provides an essential mechanism by which classical behavior emerges at macroscopic scales from quantum systems when measurement is imposed \cite{Zurek:2003zz}. In general, When a quantum system interacts with its environment, the reduced density matrix of the system from tracing out the environment appears to be mixed and classical, while decoherence manifests at the loss of off-diagonal terms in the reduced density matrix. However, it should be noted that decoherence does not violate entanglement. When decoherence occurs, the entanglement in the initial quantum system is transferred to the system-environment entanglement.

Gravitational decoherence, referring to the loss of coherence in quantum superpositions, is related to gravitational effects arising from a classical or quantum gravitational background \cite{Bassi:2017szd}. Classical gravitational backgrounds include primordial gravitational waves, the cosmic microwave background radiation, stochastic gravitational waves, and others. Recently, Danielson, Satishchandran, and Wald (DSW) investigated a Gedankenexperiment \cite{Danielson:2022tdw}, where the spatial superposition is treated quantum mechanically while the background gravitational field is a classical black hole. They demonstrated that the presence of black hole killing horizon can inevitably decohere the quantum superposition of a charged or massive particle \cite{Danielson:2022tdw,Danielson:2022sga}.

The setup of the DSW Gedankenexperiment is described as follows \cite{Danielson:2022tdw}. Suppose that in a static black hole spacetime, a charged or massive particle is initially prepared in the positive $x$-spin state. A local observer sends the particle through a Stern-Gerlach apparatus oriented in the $z-$ direction. After that, the particle is in a superposition state of the following form
\begin{eqnarray}\label{superposition_state}
    |\Psi\rangle= \frac{1}{\sqrt{2}}\left(|\psi_L,\uparrow\rangle+|\psi_R,\downarrow\rangle\right)\;,
\end{eqnarray}
where $|\psi_L\rangle$ and $|\psi_R\rangle$ are the spatially separated, normalized states of the particle after passing through the apparatus and $|\uparrow\rangle$ and $|\downarrow\rangle$ are the spin-up and spin-down states along the $z-$axis direction. The local observer maintains the stationary superposition state of the particle for a duration of $T$, and then recombines the particle using a reverse Stern-Gerlach apparatus. After recombination, the particle can, in principle, be measured to verify whether the initial state has been restored. If the coherence of the superposition state \eqref{superposition_state} is preserved over a long period of time $T$, the particle's spin will consistently be found in the initial positive $x$-direction.

Assuming that the separation and recombination processes occur adiabatically, and that no external influences affect the particle that could induce potential decoherence, analyses have shown that coherence can be maintained in Minkowski spacetime \cite{Belenchia:2018szb, Danielson:2021egj}. However, an analysis of the Gedankenexperiment in black hole spacetime leads to a dramatically different conclusion \cite{Danielson:2022tdw, Danielson:2022sga}. It has been shown that the low-frequency electromagnetic or gravitational radiation emitted by the charged or massive particle inevitably interacts with the black hole horizon, effectively disappearing from the outside spacetime due to the causal structure of the horizon. This radiation may provide ``which-way" information about the particle, analogous to an observer behind the horizon performing a ``which-way" experiment. Consequently, the coherence of the superposition state of the charged or massive particle is inevitably violated. The decoherence of a quantum superposition induced by a black hole is referred to as DSW decoherence.

The effect of DSW decoherence has attracted significant attention. In \cite{Gralla:2023oya}, the decoherence of quantum superposition in the rotating Kerr black hole was studied, with decoherence rates calculated for the scalar and electromagnetic cases. However, the gravitational case was not addressed. Along similar lines, we also examine the decoherence of quantum superposition in the charged Reissner-Nordström black hole \cite{Li:2024guo}. In these studies, the expected number of entangling photons or gravitons emitted by the experimental particle—related to the decoherence rate—was evaluated at the past or future horizon. This framework is referred to as the global description.

More recently, it has been recognized that DSW decoherence should be interpreted from the perspective of the local observer conducting the superposition experiment \cite{Wilson-Gerow:2024ljx, Biggs:2024dgp, Danielson:2024yru}. In other words, the experimenter must also have a local description for the cause of decoherence. In \cite{Wilson-Gerow:2024ljx}, a local description was proposed by mapping the DSW setup onto a worldline-localized model resembling an accelerating Unruh-DeWitt particle detector in Minkowski spacetime. It was shown that the thermal environment experienced by the accelerating observer due to the Unruh effect \cite{Unruh:1976db} induces Ohmic friction on the experimental system, which is the key mechanism behind steady decoherence as implied by the fluctuation-dissipation theorem. In \cite{Biggs:2024dgp}, by modeling black hole as a quantum system at finite temperature, it was also shown that the decoherence phenomenon could also be interpreted in this context. Furthermore, it was analyzed that the decoherence effect caused by the black hole horizon is comparable to the analogous effect induced by the ordinary matter system. The two previous papers indicated that decoherence is related to the local two-point function of the quantum field within the local observer's lab. This idea was further developed in \cite{Danielson:2024yru}, where the number of entangling particles was explicitly expressed in terms of the local two-point function. Therefore, DSW decoherence is a fascinating effect, and understanding its nature could provide valuable insights into both the quantum properties of black holes and the black hole information paradox.

In this work, we study the DSW decoherence effect for the quantum superposition in dS spacetime due to the presence of the cosmological horizon. Although this effect was previously discussed by DSW in \cite{Danielson:2022sga} from the global description perspective, the precise numerical coefficients in the expressions for the decoherence rates of quantum fields with different spins remain undetermined. We will use the algebraic approach of quantum field theory on curved spacetime \cite{Wald:1995yp, Hollands:2014eia} to derive the exact expression for the expected number of entangling particles in the scalar field case. Note that the electromagnetic and gravitational cases were addressed in \cite{Danielson:2024yru}. The expressions for the entangling particles establish a direct relation between the decoherence rate and the local two-point correlation function.

Specifically, we analyze the quantum superposition Gendankenexperiment performed by a local observer at the center of dS spacetime. For the local description of DSW decoherence in dS spacetime, we work in the dS conformal invariant vacuum. The two-point correlation functions for the conformally coupled scalar field, the electromagnetic field, and the linearized gravitational field are extensively investigated. By using the two-point functions for the scalar \cite{Birrell:1982ix,Polarski:1989bv}, electromagnetic \cite{Allen:1985wd,Youssef:2010dw}, and linearized gravitational fields \cite{Kouris:2001hz,Mora:2012kr} in dS spacetime, we compute the corresponding entangling particle numbers. It is shown that these particle numbers are proportional to the duration for which the local observer holds the superposition state. This  demonstrates that the quantum spatial superposition is decohered by the emission of entangling particles into the cosmological horizon. Note that our setup is equivalent to that of an accelerating observer in 5-dimensional Minkowski spacetime. The results for the scalar and electromagnetic cases are consistent with those obtained in \cite{Wilson-Gerow:2024ljx}, which studied the decoherence effect from the perspective of an accelerating observer in Minkowski spacetime. However, our result fixes the numerical coefficient of the gravitational decoherence rate for the first time.

This paper is organized as follows. In Sec.\ref{sec:decoh_ent_scalar}, we derive an explicit expression for the entangling particle number in the scalar field case using the algebraic approach of quantum field theory on curved spacetime. This expression establishes the relationship between the entangling particle number and the local two-point correlation function of the scalar field. In Sec.\ref{Sec:Dec_dS_scalar}, we calculate the entangling particle number for the scalar field in dS spacetime, employing the correlation function for the scalar field in the dS-invariant vacuum. In Sec.\ref{sec:Dec_ent_photon}, we extend the calculation to the electromagnetic field case and determine the expected number of entangling photons. In Sec.\ref{Sec:Dec_ent_graviton}, we investigate the decoherence effect for the linearized gravitational field in dS spacetime. The conclusion and discussion are presented in the final section.

\section{Local expression of decoherence from scalar field radiation}
\label{sec:decoh_ent_scalar}

In this section, we derive an explicit expression for the entangling particle number in the scalar field case using the algebraic approach of quantum field theory on curved spacetime.

For the massless minimally coupled scalar field in dS spacetime, there exists no conformally invariant vacuum state \cite{Allen:1985ux}. For the massive scalar fields, such a vacuum state does exist. For this reason, we consider a massive conformally coupled scalar field $\phi$ in a curved spacetime $\mathcal{M}$. The following derivation also applies to general spacetime $\mathcal{M}$ with a static Killing vector, with the static patch of dS spacetime as a special example.

The dynamics of the conformally coupled scalar field is governed by the following Klein-Gordon equation  
\begin{eqnarray}\label{K_G_eq}
    \left( \Box-m^2- \xi R \right) \phi(x)=0\;,
\end{eqnarray}
where $\Box$ denotes the d'Alembert operator in curved spacetime, $m$ is the mass of the scalar field, $\xi$ is the coupling constant, and $R$ is the scalar curvature of $\mathcal{M}$. For conformal coupling, $\xi=\frac{1}{6}$.

In the algebraic approach of quantum field theory (see \cite{Wald:1995yp} and \cite{Hollands:2014eia} for the nice intoduction and review), the quantized scalar field operator $\hat{\phi}$ is typically defined as the "smeared" field to resolve the singularity in the correlation function when two spacetime points coincide. For a classical solution $\phi(x)$ to the equation of motion, the "smeared" field operator is defined as
\begin{eqnarray}\label{Smearing_def}
    \hat{\phi}(f)=\int\sqrt{-g} d^4 x f(x) \phi(x) \;,
\end{eqnarray}
where $f$ is a smooth test function with compact support on $\mathcal{M}$. In this sense, the field equation \eqref{K_G_eq} is equivalent to 
\begin{eqnarray}
    \hat{\phi}\left(\left( \Box-m^2- \xi R \right)f\right)&=&
    \int\sqrt{-g} d^4 x \left( \Box-m^2- \xi R \right)f(x) \phi(x) 
    \nonumber\\
   &=& \int \sqrt{-g} d^4 x  f(x) \left( \Box-m^2- \xi R \right) \phi(x) 
    \nonumber\\
    &=&0\;,
\end{eqnarray}
where the integration by parts is performed twice.

In the following, we will consider the scalar radiations emitted by the particle holding by the local observer. Therefore, the particle holding by the observer should be treated as the source of the scalar field. We will mainly focus on the Klein-Gordon equation with a source $\rho$ as follows
\begin{eqnarray}\label{Sourced_KGeq}
    \left( \Box-m^2- \xi R \right) \phi(x)=\rho(x)\;.
\end{eqnarray}
The solution to the sourced Klein-Gordon equation \eqref{Sourced_KGeq} can be obtained by adding a particular solution to the solution of the sourceless Klein-Gordon equation \eqref{K_G_eq}. As claimed in \cite{Gralla:2023oya}, the choice of the particular solution determines the state of the scalar field radiation, which is essential to derive the expression of the entangling particle number.

Prior to the experimental particle going through the Stern-Gerlach apparatus, we may consider the observable of the scalar field radiation  
\begin{eqnarray}\label{radiation_field_def}
    \hat{\phi}^{\textrm{in}}=\hat{\phi}-G^{\textrm{ret}}\hat{I}\;,
\end{eqnarray}
where $G^{\textrm{ret}}$ is the classical retarded solution to the sourced Klein-Gordon equation \eqref{Sourced_KGeq} and $\hat{I}$ denotes the identity operator. This is analogous to the electromagnetic case considered in \cite{Danielson:2022sga}, where the observable of the electromagnetic radiation is defined by subtracting the ``Coulomb" part from the electromagnetic field. It is obvious that the radiation observable $\hat{\phi}^{\textrm{in}}$ satisfies the sourceless Klein-Gordon equation \eqref{K_G_eq}.

The initial state of the scalar radiation is assumed to be in the vacuum state $|\Omega\rangle$, which in the present case is the de Sitter-invariant vacuum state for the conformally coupled scalar field. For the general case of black hole spacetime formed by gravitational collapse, the Unruh vacuum is a natural selection of the vacuum state $|\Omega\rangle$.

From the vacuum state $|\Omega\rangle$, one can construct a one-particle Hilbert space $\mathcal{H}_{\textrm{in}}$ and the corresponding Fock space $\mathcal{F}(\mathcal{H}_{\textrm{in}})$\cite{Kay:1988mu}. The field operator $\hat{\phi}^{\textrm{in}}$ is represented on $\mathcal{F}(\mathcal{H}_{\textrm{in}})$ as 
\begin{eqnarray}\label{operator_decom}
    \hat{\phi}^{\textrm{in}}(f)=i \hat{a} \left( \overline{K\Delta f}\right)-i a^\dagger(K\Delta f)\;.
\end{eqnarray}
Here $f$ is an arbitrary test function with compact support, $\Delta f$ is the advanced minus retarded solution to the Klein-Gordon equation with the source $f$, and $K$ denotes the map from the classical solution space to the one particle Hilbert space $\mathcal{H}_{\textrm{in}}$. The creation and annihilation operators satisfy the commutation relation 
\begin{eqnarray}\label{commun_relation}
    \left[\hat{a} \left( \overline{K\Delta f}\right), \hat{a}^\dagger \left( K\Delta g\right)  \right]=\langle K\Delta f| K\Delta g \rangle \hat{I}\;,
\end{eqnarray}
where the overline denotes the complex conjugate, and $\langle K\Delta f| K\Delta g \rangle$ is the inner product on $\mathcal{H}_{\textrm{in}}$. For the scalar field case, the inner product on $\mathcal{H}_{\textrm{in}}$ is just a natural generalization of the Klein-Gordon inner product.

After the particle passing through the Stern-Gerlach apparatus, the particle is in a spatially separated quantum superposition state, which effectively results in two possible evolutions of the radiation field. The state of the total particle-radiation system is then given by the following form 
\begin{eqnarray}\label{superposition_state_total}
    |\Psi\rangle= \frac{1}{\sqrt{2}}\left(|\psi_L,\uparrow\rangle\otimes |\Psi_L\rangle+|\psi_R,\downarrow\rangle\otimes |\Psi_R\rangle\right)\;,
\end{eqnarray}
where $|\Psi_L\rangle$ and $|\Psi_R\rangle$ are the ``out" states of the scalar radiation corresponding to the particle states $|\psi_L\rangle$ and $|\psi_R\rangle$, respectively. The decoherence due to the scalar radiation is then given by \cite{Danielson:2022tdw}
\begin{eqnarray}\label{Decoherence_def}
    \mathcal{D}=1-|\langle \Psi_L|\Psi_R\rangle|\;.
\end{eqnarray}
If the ``out" states of the scalar radiation are clearly distinguishable, i.e. $\langle\Psi_L|\Psi_R\rangle=0$, the quantum superposition is decohered finally. Therefore, to determine whether the quantum state \eqref{superposition_state_total} has decohered, one must calculate the overlap between $|\Psi_L\rangle$ and $|\Psi_R\rangle$ after the observer has held the experimental system for a long time $T$.

To calculate the overlap of $|\Psi_L\rangle$ and $|\Psi_R\rangle$ for the scalar radiation, it is convenient to work in the Heisenberg representation. It is assumed that the charge densities $\rho_i (i=L,R)$ corresponding to the particle state $|\psi_i\rangle$ can be treated as c-number source in Klein-Gordon equation \eqref{Sourced_KGeq}. After the particle going through the Stern-Gerlach apparatus, the scalar field operator $\hat{\phi}_{i}$ corresponding to the state $|\psi_i\rangle$ can be expressed as 
\begin{eqnarray}
    \hat{\phi}_{i}=\hat{\phi}^{\textrm{in}}+G^{\textrm{ret}}(\rho_i) \hat{I}\;,
\end{eqnarray}
where $G^{\textrm{ret}}(\rho_i)$ is the classical retarded solution to Klein-Gordon equation with the source $\rho_i$.

After the time at which the particle is recombined, we have $\rho=\rho_L=\rho_R$. In order to describe the observable of the ``out" radiation field, the scalar field operator should be expressed as \cite{Gralla:2023oya}
\begin{eqnarray}
    \hat{\phi}_{i}=\hat{\phi}_i^{\textrm{out}}+G^{\textrm{adv}}(\rho_i) \hat{I}\;,
\end{eqnarray}
where $G^{\textrm{adv}}(\rho_i)$ is the classical retarded solution to Klein-Gordon equation with the source $\rho_i$. 

It is easy to obtain the relations between the ``in" and ``out" radiation fields as 
\begin{eqnarray}
    \hat{\phi}_L^{\textrm{out}}=\hat{\phi}^{\textrm{in}}+\left[G^{\textrm{ret}}(\rho_L)-G^{\textrm{adv}}(\rho_L)  \right] \hat{I}\;,\nonumber\\
    \hat{\phi}_R^{\textrm{out}}=\hat{\phi}^{\textrm{in}}+\left[G^{\textrm{ret}}(\rho_R)-G^{\textrm{adv}}(\rho_R)  \right] \hat{I}\;.
\end{eqnarray}
From the general theory of coherent state \cite{Glauber:1963tx,Zhang:1990fy,2012JPhA...45x4002S}, these two equations imply that the ``out'' states of the radiation fields corresponding to the ``in'' vacuum state $|\Omega\rangle$ are given by the coherent states with the following relation 
\begin{eqnarray}
    |\Psi_i\rangle = \textrm{e}^{-\frac{1}{2}\big\| K G(\rho_i)\big\|^2 } \exp \left[ \hat{a}^\dagger \left( K G(\rho_i) \right) \right]|\Omega\rangle\;,
\end{eqnarray}
where $ G(\rho_i)=G^{\textrm{ret}}(\rho_i)-G^{\textrm{adv}}(\rho_i)$ is the difference of the retarded and the advanced solution associated with the source $\rho_i$. The ``out'' states of the radiation fields are related with the ``in" vacuum state by a displacement operator.

Because the difference between the two components of the particle in the superposition, the left and the right evolutions of the radiation fields are different. In general, the overlap of the radiation states $|\Psi_L\rangle$ and $|\Psi_R\rangle$ is not unity. By using the general formula to calculate the overlap of the coherent state, one can get
\begin{eqnarray}
    |\langle \Psi_L|\Psi_R\rangle|
    =\exp\left[ -\frac{1}{2} \langle N\rangle \right]\;,
\end{eqnarray}
where
\begin{eqnarray}\label{Entangleing_photon}
    \langle N\rangle =\big\| K \Delta(\rho_R-\rho_L)\big\|^2\;,
\end{eqnarray}
with
\begin{eqnarray}
    \Delta(\rho_R-\rho_L)&=& G(\rho_R)-G(\rho_L) \nonumber\\
    &=& 
   \left[G^{\textrm{ret}}(\rho_R)-G^{\textrm{adv}}(\rho_R) \right]-\left[G^{\textrm{ret}}(\rho_L)-G^{\textrm{adv}}(\rho_L) \right]
   \nonumber\\
   &=& 
   \left[G^{\textrm{ret}}(\rho_R-\rho_L)-G^{\textrm{adv}}(\rho_R-\rho_L) \right]\;. 
\end{eqnarray}

It is clear that $\Delta (\rho_R-\rho_L)$ is the advanced minus retarded solution to the Klein-Gordon equation with the source $\rho_R-\rho_L$ and $K \Delta(\rho_R-\rho_L)$ represents the one-particle state in Hilebert space $\mathcal{H}_{\textrm{in}}$. Following DSW, $\langle N\rangle$ is referred to as the expected number of entangling particles. It is related to the decoherence of the particle state by the following relation
\begin{eqnarray}
    \mathcal{D}=1-\exp\left[-\frac{1}{2} \langle N\rangle \right]\;.
\end{eqnarray}
If $\langle N\rangle$ is significantly larger than $1$, the superposition state \eqref{superposition_state_total} will be completely decohered. For the black hole case, the decoherence arises from the entangling particles emitted by a charged or massive experimental particle. The entangling particles effectively transfer quantum information of the experimental particle into the black hole horizon, which is causally disconnected from the outside spacetime. This, in turn, creates entanglement between the black hole and the experimental particle, leading to the decoherence of the quantum spatial superposition. This physical mechanism is also applied to cosmological horizon in dS spacetime.

The main task left is to calculate the entangling particle number $\langle N\rangle$. By using the commutation relation \eqref{commun_relation}, one can get the following useful identity about the norm of the quantum state corresponding to the classical solution $\Delta(\rho_R-\rho_L)$
\begin{eqnarray}
    \big\| K \Delta(\rho_R-\rho_L)\big\|^2
    =\langle \Omega |\hat{a}\left(\overline{K\Delta(\rho_R-\rho_L)}\right)\hat{a}^\dagger\left(K\Delta(\rho_R-\rho_L)\right) 
    |\Omega\rangle\;,
\end{eqnarray}
where we have applied the fact that $|\Omega\rangle$ is the vacuum state and is annihilated by the operator $\hat{a}\left(\overline{K\Delta(\rho_R-\rho_L)}\right)$. 

On the other hand, using the Fock space representation \eqref{operator_decom} of the operator $\hat{\phi}^{\textrm{in}}$, one can get
\begin{eqnarray}
  \langle \Omega |\left[\hat{\phi}^{\textrm{in}}\left(\rho_R-\rho_L\right)\right]^2 |\Omega\rangle=  \langle \Omega |\hat{a}\left(\overline{K\Delta(\rho_R-\rho_L)}\right)\; \hat{a}^\dagger\left(K\Delta(\rho_R-\rho_L)\right) 
    |\Omega\rangle\;,
\end{eqnarray}
where $\hat{\phi}^{\textrm{in}}\left(\rho_R-\rho_L\right)$ denotes the field operator $\hat{\phi}^{\textrm{in}}$ smearing with the function $\rho_R-\rho_L$. Therefore, the entangling particle number can also be expressed as 
\begin{eqnarray}
   \langle N\rangle= \big\| K \Delta(\rho_R-\rho_L)\big\|^2
    =\langle \Omega |\left[\hat{\phi}^{\textrm{in}}\left(\rho_R-\rho_L\right)\right]^2 |\Omega\rangle\;.
\end{eqnarray}
This equation gives the local reformulation of the decoherence of quantum superposition due to the scalar radiation. Note that the electromagnetic and gravitational cases were addressed in \cite{Danielson:2024yru}. Now we want to address the relation between the entangling particle number $\langle N\rangle$ and the two-point correlation function of the scalar field.

To proceed, we neglect the spatial extent of the experimental particle and approximate the source as the point-like. For a point source described by a charge $q(\tau)$ on a worldline $x^\mu(\tau)$, with $\tau$ being the proper time parameter, the charge density can be given by 
\begin{eqnarray}
    \rho=\int  \frac{q(\tau)}{\sqrt{-g}} \delta^{(4)} \left(x-x(\tau)\right) d\tau\;,
\end{eqnarray}
where $\delta^{(4)} \left(x-x(\tau)\right)$ is the ``coordinate delta function. For the static patch of dS spacetime, there is a static killing vector $t^a=\left(\frac{\partial}{\partial t}\right)^a$. One can perform the integration with respect to the killing coordinate time $t$, which gives us the following expression for the charge density
\begin{eqnarray}
     \rho(t,x^i)=\frac{q}{\sqrt{-g}} \delta^{(3)} \left(x^i-x^i(t)\right) \frac{d\tau}{dt}\;,
\end{eqnarray}
where $x^i$ is the spatial coordinates on the hypersurfaces $\Sigma_{t}$ orthogonal to the timelike Killing vector $t^a$. For simplicity, we set the scalar charge $q(\tau)$ to be a constant $q$. Since the static patch of dS spacetime is similar to the static black hole spacetime, the following discussion can also be applied to the later case.

We denote the displacement between the two components of the experimental particle at time $t$ by the tangent vector $S^a(t)$ to the geodesic segment in $\Sigma_{t}$ of unit affine parameter that connect the centers of the two components. By introducing the unit vector $s^a$ that is Lie transported along $t^a$, $S^a(t)$ can be rewritten as $S^a=d(t)s^a$, with $d(t)$ is the proper distance between the two components. With this geometric picture, the difference of the scalar charge densities between the two components is given by
\begin{eqnarray}
    \rho_R-\rho_L&\approx&\frac{q}{\sqrt{-g}}\left[-\partial_a \delta^{(3)}\left(x^i-x^i(t)\right)  \frac{d\tau}{dt} +\delta^{(3)}\left(x^i-x^i(t)\right)\partial_a\left(\frac{d\tau}{dt}\right) \right] s^a d(t) \nonumber\\&=&    
    q \left\{-\nabla_a\left[\frac{1}{\sqrt{-g}}\delta^{(3)}\left(x^i-x^i(t)\right)\frac{d\tau}{dt}\right]+\partial_a\left(\frac{1}{\sqrt{-g}}\right)\delta^{(3)}\left(x^i-x^i(t)\right)\frac{d\tau}{dt}\right.\nonumber\\
    &&+\left.\frac{2}{\sqrt{-g}}\partial_a\left(\frac{d\tau}{dt}\right) \delta^{(3)}\left(x^i-x^i(t)\right)
    \right\} s^a d(t)\;,
\end{eqnarray}
where $x^i(t)$ can be treated as the coordinates of the local observer's lab. We have used the fact that $\frac{1}{\sqrt{-g}} \delta^{(3)} \left(x^i-x^i(t)\right) \frac{d\tau}{dt}$ is scalar function and thus 
\begin{eqnarray}
    \partial_a\left[ \frac{1}{\sqrt{-g}} \delta^{(3)} \left(x^i-x^i(t)\right) \frac{d\tau}{dt} \right]=\nabla_a\left[ \frac{1}{\sqrt{-g}} \delta^{(3)} \left(x^i-x^i(t)\right) \frac{d\tau}{dt} \right]\;.
\end{eqnarray}

The field operator $\hat{\phi}^{\textrm{in}}$ smeared in $\rho_R-\rho_L$ can be given by 
\begin{eqnarray}
    \hat{\phi}^{\textrm{in}}\left(\rho_R-\rho_L\right)
    &=&\int d^4 x \sqrt{-g}  \left(\rho_R-\rho_L\right)  \hat{\phi}^{\textrm{in}}(t,x^i) \nonumber\\
    &\approx&
    q\int d^4 x \sqrt{-g} \left\{
    -\nabla_a\left[\frac{1}{\sqrt{-g}}\delta^{(3)}\left(x^i-x^i(t)\right)\frac{d\tau}{dt} s^a d(t) \hat{\phi}^{\textrm{in}}(t,x^i)\right]\right.\nonumber\\
    &&~~~~~~~~~~~~~~~~~~~~~~~+\frac{1}{\sqrt{-g}}\delta^{(3)}\left(x^i-x^i(t)\right)\frac{d\tau}{dt} \nabla_a \left(s^a d(t) \hat{\phi}^{\textrm{in}}(t,x^i)\right) \nonumber\\
    &&~~~~~~~~~~~~~~~~~~~~~~~+\partial_a\left(\frac{1}{\sqrt{-g}}\right)\delta^{(3)}\left(x^i-x^i(t)\right)\frac{d\tau}{dt} s^a d(t) \hat{\phi}^{\textrm{in}}(t,x^i) \nonumber\\
    &&~~~~~~~~~~~~~~~~~~~~~~~+\frac{2}{\sqrt{-g}}\partial_a\left(\frac{d\tau}{dt}\right) \delta^{(3)}\left(x^i-x^i(t)\right) s^a d(t) \hat{\phi}^{\textrm{in}}(t,x^i)   
     \;.
\end{eqnarray}
By using Gauss’ theorem, the first term in the integral can be converted into a boundary integral, which can be safely dropped due to the presence of the delta function and the fact that $d(t)$ tends to zero as $t\rightarrow \pm\infty$. In addition, since $t^a$ is orthogonal to the hypersurface $\Sigma_t$, we have $s^a \nabla_a d(t)=0$. By performing the integral with respect to the delta function, we can obtain the compact expression as  
\begin{eqnarray}\label{phi_delta_rho}
    \hat{\phi}^{\textrm{in}}\left(\rho_R-\rho_L\right)
    &\approx&q\int d\tau d(\tau) \left\{ s^a\nabla_a \hat{\phi}^{\textrm{in}}(\tau,x^i) + \left(\frac{dt}{d\tau}\right)^2 \partial_a \left[s^a \left(\frac{d\tau}{dt}\right)^2\right] \hat{\phi}^{\textrm{in}}(\tau,x^i)\right\} \;.
\end{eqnarray}
In the static black hole spacetime, if the motion of the lab is non-relativistic relative to the rest frame of $t^a$ and the two components of the experimental particle are separated along the radial direction, a simple estimation shows that the second term can be neglected as long as the observer's lab is far from the horizon. In our case, we assume the lab is located at the center of the dS spacetime. In the next section, we will demonstrate that the second term does not contribute to the final result.

Therefore, the entangling particle number is then given by 
\begin{eqnarray}
    \langle N\rangle = q^2 \int d\tau d\tau' d(\tau) d(\tau')  
    s^a s^b \langle \nabla_a \hat{\phi}^{\textrm{in}}(\tau,x^i) \nabla_b \hat{\phi}^{\textrm{in}}(\tau',x'^{i}) \rangle \;.
\end{eqnarray}
This equation reproduces the relation between the entangling particle number and the local two-point correlation function of the scalar field as derived in \cite{Wilson-Gerow:2024ljx}, where the formula for the decoherence functional is obtained from the influence functional description of an open quantum system. The explicit form of the decoherence functional depends on the interaction Hamiltonian, making their formula model-dependent. In contrast, our result is clearly model-independent.

Note that the electromagnetic and gravitational cases were addressed in \cite{Danielson:2024yru}. The expressions for the entangling particles establish a direct relation between the decoherence rate and the local two-point correlation function.

\section{Decoherence rate from scalar field in de Sitter spacetime} 
\label{Sec:Dec_dS_scalar}

In this section, we will examine the quantum spatial superposition held by the local observer at the center of dS spacetime and calculate the corresponding decoherence rate due to the emission of scalar particles.

It is well known that dS spacetime can be viewed as a four dimensional hyperboloid 
\begin{eqnarray}
    -\left(x^0\right)^2+\left(x^1\right)^2+\left(x^2\right)^2+\left(x^3\right)^2+\left(x^4\right)^2=L^2\;,
\end{eqnarray}
embedded in the five dimensional Minkovski spacetime. By introducing the following coordinate transformation 
\begin{eqnarray}\label{coord_trans}
    x^0&=&\sqrt{L^2 -r^2} \sinh \frac{t}{L}\;,\nonumber\\
    x^1&=&\sqrt{L^2 -r^2} \cosh \frac{t}{L}\;,\nonumber\\
    x^2&=&r\cos \theta\;,\\
    x^3&=&r\sin\theta\cos\varphi\;,\nonumber\\
    x^4&=&r\sin\theta\sin\varphi\;,\nonumber
\end{eqnarray}
the static dS metric can be expressed as  
\begin{eqnarray}\label{dS_metric}
    ds^2=-\left( 1-\frac{r^2}{L^2} \right) dt^2+\frac{dr^2}{\left( 1-\frac{r^2}{L^2} \right)}+r^2 d\Omega^2\;,
\end{eqnarray}
where $d\Omega^2$ is the line element on $S^2$. 

For the static patch of dS spacetime, there is a Killing horizon $r=L$ of the static Killing field $\left(\frac{\partial}{\partial t}\right)^a$, which is also called cosmological horizon. The cosmological horizon shares similar properties with the event horizon of black hole \cite{Gibbons:1977mu}. The Hawking radiation effect for a massless conformally coupled scalar field in static dS spacetime was studied in \cite{Polarski:1989bv}. As a result, the cosmological horizon has a Hawking temperature given by $T_H=\frac{1}{2\pi L}$. 

With the geometry of static dS spacetime given by \eqref{dS_metric}, we now explain why the second term in Eq.\eqref{phi_delta_rho} can be neglected for a non-relativistic observer at the center $r=0$ of dS spacetime. For such a observer, we have $\frac{d\tau}{dt}\approx\sqrt{-g_{tt}}=\sqrt{1-\frac{r^2}{L^2}}$. If the two components of the experimental particle are separated along the $r$ direction, the non-vanishing component of the unit vector $s^a$ is $s^r=\sqrt{1-\frac{r^2}{L^2}}$. Thus we have 
\begin{eqnarray}
    \left(\frac{dt}{d\tau}\right)^2 \partial_a \left[s^a \left(\frac{d\tau}{dt}\right)^2\right]=-\frac{3r/L^2}{\sqrt{1-\frac{r^2}{L^2}}}\;,
\end{eqnarray}
which clearly vanishes at $r=0$. Therefore, in our setup, the second term in Eq.\eqref{phi_delta_rho} can be appropriately neglected. 

For our purpose, we consider the massless conformally coupled scalar field. As claimed in the previous section, we choose the de Sitter-invariant vacuum state as vacuum state of the ``in" scalar field radiation $\phi^{\textrm{in}}$. Then the two point correlation function can be written as \cite{Birrell:1982ix,Polarski:1989bv}
\begin{eqnarray}
    \langle \phi^{\textrm{in}}(x) \phi^{\textrm{in}}(x') \rangle
    =\frac{1}{4\pi^2} \frac{1}{\Delta^2(x-x')}\;,
\end{eqnarray}
where $\Delta^2(x-x')=-\left(x^0-x'^0 \right)^2+\left(x^1-x'^1 \right)^2+\left(x^2-x'^2 \right)^2+\left(x^3-x'^3 \right)^2+\left(x^4-x'^4 \right)^2$ is the spacetime interval between the two points $x$ and $x'$. Here, $x$ and $x'$ are the coordinates of two points embedded in the 5-dimensional Minkowski spacetime.

We consider the case that the locations of the two point-like sources are separated along the $r$ direction, i.e. the coordinates of the sources are given by $\left(t,r,\theta,\varphi \right)$ and $\left(t',r',\theta,\varphi \right)$. By using the coordinate transformation \eqref{coord_trans}, We can obtain the correlation function as 
\begin{eqnarray}
    \langle \phi^{\textrm{in}}(t,r,\theta,\varphi) \phi^{\textrm{in}}(t',r',\theta,\varphi) \rangle
    =-\frac{1}{8\pi^2}\left[rr'-L^2+\sqrt{L^2-r^2}\sqrt{L^2-r'^2}\cosh\left(t-t'\right)/L\right]^{-1}\;.
\end{eqnarray}

Furthermore, we consider the case that when the pointlike source pass through the Stern-Gerlach apparatus, the two components of the pointlike source are separated from the point $r=0$ with $r+r'=0$ and $\Delta r=r'-r\approx d\ll 1$ for a very long proper time $T$. In the reference of the local observer's lab, i.e. $r=0$, $T$ is the proper time of the local observer and the $d(t)$ is maintained constant $d$ for a long time. In this case, one can get 
\begin{eqnarray}\label{correl_func_rr}
    \langle \partial_r \hat{\phi}^{\textrm{in}} \partial_{r'} \hat{\phi}^{\textrm{in}} \rangle \approx 
    \frac{1}{32\pi^2 L^4} \frac{1}{\sinh^{4} \left(\tau-\tau'\right)/2L}\;,
\end{eqnarray}
where we have used the approximation condition $r=r'\approx 0$. Note that in this case $t-t'=\tau-\tau'$ when $r=r'\approx 0$. Therefore, the correlation function \eqref{correl_func_rr} is just a function of $\Delta\tau=\tau-\tau'$ finally.

For the components of the particle separated along the $r$-direction, we have 
\begin{eqnarray}
    s^a \nabla_a \hat{\phi}^{\textrm{in}}=s^r \partial_r \hat{\phi}^{\textrm{in}}\approx \partial_r \hat{\phi}^{\textrm{in}}\;,
\end{eqnarray}
where $s^r=\sqrt{1-\frac{r^2}{L^2}}$ is the unit displacement vector along the $r$ direction. We also used the approximation condition that near $r=0$, $s^r\approx 1$. The entangling particle number can be given by 
\begin{eqnarray}
    \langle N\rangle \approx q^2 \int d\tau d\tau' d(\tau)d(\tau')\langle \partial_r \hat{\phi}^{\textrm{in}} \partial_{r'} \hat{\phi}^{\textrm{in}} \rangle\;. 
\end{eqnarray}

We assume that $d(\tau)$ is a function as 
\begin{eqnarray}
    d(\tau)=\begin{cases}
d, & |\tau|<T/2\;,\\
0, & \tau<-T/2-T_1 \;\;\textrm{or}\;\; \tau>T/2+T_2\;,
\end{cases}
\end{eqnarray}
where $T$ is the proper time that the experimenter holds the superposition state and $T_1$ and $T_2$ are the time that the times that used to separate and recombine the superposition. One can roughly approximate the function $d(t)$ as a rectangular wave function to estimate the entangling particle number. 
By using 
\begin{eqnarray}
    d(\tau)=\int \frac{d\omega}{2\pi} \tilde{d}(\omega) e^{-i\omega \tau}\;,
\end{eqnarray}
one can get 
\begin{eqnarray}
    \langle N\rangle &\approx& q^2 \int \frac{d\omega}{2\pi}  \tilde{d}(\omega)\tilde{d}^*(\omega) \mathcal{F}(\omega)\;,
\end{eqnarray}
where $\tilde{d}(\omega)$ is the Fourier transform of $d(t)$ and $\mathcal{F}(\omega)$ is the Fourier transform of the correlation function $\langle \partial_r \hat{\phi}^{\textrm{in}} \partial_{r'} \hat{\phi}^{\textrm{in}} \rangle$. Since $d(t)$ can be approximated as a rectangular wave function, for large $T$, $\tilde{d}(\omega)$ is highly bandlimited near $\omega\sim 0$.

The Fourier transform of the correlation function \eqref{correl_func_rr} can be caluclated as 
\begin{eqnarray}
    \mathcal{F}(\omega)
    &=& \int_{-\infty}^{+\infty} \langle \partial_r \hat{\phi}^{\textrm{in}} \partial_{r'} \hat{\phi}^{\textrm{in}} \rangle e^{i\omega \Delta\tau} d\Delta\tau\;\nonumber\\
    &=&\frac{1}{32\pi^2 L^4} \int_{-\infty}^{+\infty} \frac{1}{\sinh^4 \Delta\tau/2L} e^{i\omega \Delta\tau} d\Delta\tau\;\nonumber\\
    &=&\frac{1}{16\pi^2 L^3} \times 2\pi i \times \sum_{n=0}^{+\infty}\left[\left(2iL\omega\right)^3-4 \left(2iL\omega\right)\right] e^{-2n\pi L \omega}
    \nonumber\\
    &=&\frac{1}{6\pi L^2} \cdot\frac{\omega\left(L^2 \omega^2+1\right)}{1-e^{-2\pi L \omega}}\;.
\end{eqnarray}
Therefore, we have
\begin{eqnarray}
    \langle N\rangle &\approx& q^2 \mathcal{F}(\omega=0) \int \frac{d\omega}{2\pi}  \tilde{d}(\omega)\tilde{d}^*(\omega)\nonumber\\
    &\approx&  q^2 \mathcal{F}(\omega=0) \int d\tau d^2(\tau) \nonumber\\ 
    &\approx& \frac{q^2 d^2 T}{12\pi^2 L^3}\;.
\end{eqnarray}
Note that, in the second line, we have used the Parseval theorem to convert the integral in the frequency domain to the time domain.

This result shows that the number of entangled particles emitted by the scalar source increases constantly with respect to the lab's proper time. This implies that the quantum spatial superposition decoheres at a constant rate
\begin{eqnarray}\label{dec_rate}
    \Gamma_s=\frac{q^2 d^2 }{12\pi^2 L^3}\;.
\end{eqnarray}

With the coordinate transformation given in Eq.\eqref{coord_trans}, we observe that a local observer at the center of a 4-dimensional dS spacetime is equivalent to an accelerating observer in 5-dimensional Minkowski spacetime. The result for the decoherence rate, given in Eq. \eqref{dec_rate}, is consistent with the findings in \cite{Wilson-Gerow:2024ljx}, where a scalar dipole model is employed to study the decoherence of the quantum superposition. As claimed in the previous section, the result obtained in \cite{Wilson-Gerow:2024ljx} is model-dependent, whereas our result is model-independent.

\section{Decoherence rate from electromagnetic field in de Sitter spacetime}
\label{sec:Dec_ent_photon}

In this section, we will examine the quantum spatial superposition held by the local observer at the center of dS spacetime and calculate the corresponding decoherence rate due to the emission of entangling photons.

Suppose now that the experimental particle is charged, and the decoherence of quantum spatial superposition is caused by emitting entangling photons. From the local description of the decoherence \cite{Danielson:2024yru}, the expected number of the entangling photons are given by the two-point correlation function of the electric field as 
\begin{eqnarray}\label{entangle_photon}
   \langle N \rangle&\approx& \int d\tau d\tau' d(\tau) d(\tau') \langle s^a E_{a}^{\textrm{in}}(\tau,x^i) s^{a'} E_{a'}^{\textrm{in}}(\tau',x'^i)\rangle\nonumber\\
   &=& \int d\tau d\tau' d(\tau) d(\tau') \langle 
   F_{t r} F_{t r'}
   \rangle\;,
\end{eqnarray}
where the electric field $E_a$ on a static hypersurface $\Sigma_t$ is defined as $E_a=F_{ab} t^b$ with $F_{ab}$ being the electromagnetic field tensor. For the case that the two components of the experimental particle are separated along the radial direction, we just need to calculate the two-point correlation function of the radial component of $E_{a}^{\textrm{in}}$.

The massive and massless vector two-point functions in maximally symmetric spaces of any number of dimensions have been calculated in \cite{Allen:1985wd}. The electromagnetic field correlation function in dS spacetime for the conformally invariant vacuum is given by \cite{Allen:1985wd} (see also \cite{Youssef:2010dw})
\begin{eqnarray}\label{electric_tpf}
    \langle F_{ab} F_{a'b'} \rangle =\frac{1}{8\pi^2 L^4} \frac{1}{(1-z)^2} \left[ g_{*[a[a'} g_{b]b']*} +4 n_{[a} g_{b][b'} n_{a'} \right]\;,
\end{eqnarray}
where $\ast[$, $]\ast$ are used to denote the open-bracket and close-bracket respectively of the second antisymmetrisation. Here $z=\cos^2\left(\frac{\mu}{2L}\right)$ with $\mu(x,x')$ denotes the geodesic distance between the points $x$ and $x'$ in dS space. Therefore, $\mu^2>0$ for spacelike separation and $\mu^2<0$ for timelike separation. Especially, for the timelike separation, $z=\cos\left(\frac{i(\tau-\tau')}{2L}\right)=\cosh^2\frac{(\tau-\tau')}{2L}$, where $(\tau-\tau')$ is the geodesic proper time difference.

Note that $g_{aa'}(x,x')$ is defined as the parallel propagator via the following relations 
\begin{eqnarray}
    g_{ab}(x)=g_{a}^{\;\;\;c'}(x,x')g_{c'b}(x,x')\;,\nonumber
    \\
    g_{a'b'}(x')=g_{a'}^{\;\;\;c}(x,x')g_{cb'}(x,x')\;.
\end{eqnarray}
By introducing $n_a=\nabla_a \mu$ and $n_{a'}=\nabla_{a'} \mu$, which are the unit tangents at points $x$ and $x'$, respectively, it can be proved that the parallel propagator is explicitly given by \cite{Allen:1985wd}
\begin{eqnarray}
    g_{aa'}=C(\mu)^{-1} \nabla_a n_{b'}-n_a n_{b'}\;,
\end{eqnarray}
where $C(\mu)=-\frac{1}{L} \csc\frac{\mu}{L}$.

As discussed in the previous section, we consider the two pointlike sources with the coordinates $\left(t,r,\theta,\varphi \right)$ and $\left(t',r',\theta,\varphi \right)$. The geodesic distance between two pointlike source can be calculated by embedding dS spacetime into Minkovski spacetime using the coordinate transformation \eqref{coord_trans}. It is well known that there is a simple relation between the spacetime interval $\Delta^2(x-x')$ in the embedded Minkovski spacetime and the geodesic distance $\mu(x,x')$ in dS spacetime \cite{Spradlin:2001pw} 
\begin{eqnarray}
    \Delta^2(x-x')=2L^2\left(1-\cos\left(\frac{\mu(x,x')}{L}\right)\right)\;.
\end{eqnarray}

Furthermore, we consider the case that when the pointlike source pass through the Stern-Gerlach apparatus, the two components of the pointlike source are separated from $r=0$ point with $r+r'=0$ and $\Delta r=r'-r\approx d\ll 1$ for a very long time $T$. In the experimenter's lab reference, i.e. $r=0$, $T$ is is the proper time and the $d(t)$ is maintained constant $d$ for a long time. This is to say that we are considering the two sources are timelike separated. In this case, one can get the novanishing components of $n_a$, $n_{a'}$, and $g_{ab'}$ as 
\begin{eqnarray}
    n_t=-n_{t'}=i\;,\;\;\;g_{tt'}=-g_{rr'}=-1\;.
\end{eqnarray}
It can be proved that $z-1\approx\sinh^2\frac{\tau-\tau'}{2L}$ for the two components separated near $r=0$.

For our case, we also consider the case that the two components are separated along the $r$ direction. We just need to evaluate the correlation function $\langle F_{t r} F_{t' r'} \rangle$. By employing the above relations, from Eq.\eqref{electric_tpf}, one can get  
\begin{eqnarray}
    \langle F_{t r} F_{t' r'} \rangle \approx 
    \frac{1}{8\pi^2 L^4}  \frac{1}{(1-z)^2} \left[ -\frac{1}{2} +\partial_t \mu \partial_{t'} \mu \right]\;.
\end{eqnarray}
Using the fact that $n_t n_{t'}=\partial_t \mu \partial_{t'} \mu=1$, one can finally get
\begin{eqnarray}
    \langle F_{t r} F_{t' r'} \rangle \approx 
    \frac{1}{16\pi^2 L^4}  \frac{1}{\sinh^4(\tau-\tau')/2L} \;.
\end{eqnarray}
This result can also be derived from considering the electromagnetic two-point functions obtained in \cite{Saharian:2013yya}.

By using the expression \eqref{entangle_photon} for the entangling photon number, and repeating the procedure in the scalar field case, one can get
\begin{eqnarray}
   \langle N \rangle= \frac{q^2d^2 T}{6\pi^2 L^3}\;,
\end{eqnarray}
which is twice as large as the scalar case. This result shows that the number of entangled photons emitted by the charged source increases constantly with respect to the lab's proper time. Therefore, quantum spatial superposition decoheres at a constant rate
\begin{eqnarray}\label{dec_rate_elec}
    \Gamma_e=\frac{q^2 d^2 }{6\pi^2 L^3}\;.
\end{eqnarray}

Once again, since our setup is equivalent to that in \cite{Wilson-Gerow:2024ljx}, the result is also consistent with the findings in \cite{Wilson-Gerow:2024ljx}, where the electric dipole model is employed. However, our result is independent of the interaction Hamiltonian. In fact, the decoherence due to the entangling photons was also discussed in \cite{Danielson:2022sga} from the global description. Our result fix the numerical coefficient for the decoherence rate.

\section{Decoherence rate from linearized gravitational field in de Sitter spacetime}
\label{Sec:Dec_ent_graviton}

In this section, we consider the case that the decoherence of quantum superposition for the experimental particle is violated by emitting entangling gravitons.

The corresponding expression for the entangling graviton number is given by \cite{Danielson:2024yru}
\begin{eqnarray}
   \langle N \rangle&=&\left\langle \Omega\left|\left[ h_{ab}^{\textrm{in}}\left(T_1^{ab}-T_2^{ab}\right) \right]^2\right|\Omega\right\rangle \nonumber\\
   &\approx& m^2 \int dt dt' d^2(t) d^2(t') s^a s^b s^c s^d 
   \left\langle  E_{ab}^{\textrm{in}}(t,x^i) E_{cd}^{\textrm{in}}(t',x'^i) \right\rangle \;,
\end{eqnarray}
where $h_{ab}$ is the linearized gravitational perturbation, 
$T_1^{ab}-T_2^{ab}$ is the difference in the energy-momentum tensor of the two components of the experimental particle, and $E_{ab}=C_{acbd}t^c t^d$ is the quantum field observable corresponding to the electric part of the Weyl tensor.

For the experimenter at the center of dS spacetime $r=0$, the time $t$ can also be changed into the proper time $\tau$. Thus the calculation of the expected number of the entangling gravitons reduces to the computation of the two-point correlation function of the Weyl tensor $C_{abcd}$.

In \cite{Higuchi:2001uv}, the covariant graviton propagator in de Sitter spacetime with one gauge parameter was calculated. Based on the result, an explicit form for the two-point function of the Weyl tensor in de Sitter spacetime was also obtained in \cite{Kouris:2001hz}. The result is presented in Appendix A. For our case, we need to calculate the two-point function $\langle C_{rtrt}(t,x^i) C_{r't'r't'} (t',x'^i) \rangle$. After some algebra, one can get 
\begin{eqnarray}\label{Weyl_TPF}
    \langle C_{rtrt}(t,x^i) C_{r't'r't'} (t',x'^i) \rangle
    =-\frac{1}{16\pi^2L^6} \frac{1}{\sinh^6(t-t')/2L}\;.
\end{eqnarray}
To get the final result, we then need to calculate the Fourier transform of the function $\sinh^{-6}(\Delta \tau)/2L$ with $\Delta\tau=\tau-\tau'\approx t-t'$. It is shown that its Fourier transform is given by 
\begin{eqnarray}
    \mathcal{F}(\sinh^{-6} \Delta\tau/2L)&=&
    \int_{-\infty}^{+\infty} \frac{1}{\sinh^6\Delta\tau/2L} e^{i\omega \Delta\tau} d\Delta\tau\nonumber\\
    &=&\frac{4\pi i L}{5!}\left[(2iL\omega)^5-20 (2iL\omega)^3 +64 (2iL\omega)\right]\frac{1}{1-e^{-2\pi L \omega}}\;,
\end{eqnarray}
where only the residues at the origin and in the upper half complex plane contribute the contour integral. According to the arguments in the previous sections, what we need is just the value of the Fourier transform at $\omega=0$. By taking the limit $\omega\rightarrow 0$, we get the limit value is $-\frac{32}{15}L$. Therefore, repeating the procedure in the scalar case, we find that the entangling graviton number is given by 
\begin{eqnarray}
    \langle N \rangle = \frac{2m^2 d^4 T}{15\pi^2 L^5}.
\end{eqnarray}
Our result is consistent with the previous result obtain by DSW in \cite{Danielson:2022sga}. However, our result fixes the numerical coefficient as $\frac{2}{15\pi^2}$. 

This result also shows that the number of entangled gravitons emitted by the massive source increases constantly with respect to the lab's proper time. Therefore, quantum spatial superposition decoheres at a constant rate
\begin{eqnarray}\label{dec_rate_grav}
    \Gamma_g=\frac{2m^2 d^4 }{15\pi^2 L^5}\;.
\end{eqnarray}

\section{Conclusion and discussion}
\label{sec:con_disc}

In summary, we have studied the DSW decoherence effect for the quantum superposition in dS spacetime due to the presence of the cosmological horizon. By employing the algebraic approach of quantum field theory on curved spacetime, we derive the exact expression for the expected number of entangling particles in the scalar field case. This result together with the results for the electromagnetic and gravitational cases establish a direct relation between the decoherence and the local two-point correlation function.

The quantum superposition Gendankenexperiment performed by a local observer at the center of dS spacetime was analyzed in detail. Specifically, we compute the decoherence rate caused by emission of entangling scalar particles, photons, and gravitons, respectively. It is shown that quantum spatial superposition decoheres at a constant rate with respect to the local observer's proper time. Our setup is equivalent to that of an accelerating observer in 5-dimensional Minkowski spacetime. The results for the scalar and electromagnetic cases are consistent with those obtained in \cite{Wilson-Gerow:2024ljx}. The discussions in \cite{Wilson-Gerow:2024ljx} is model-dependent, whereas our result is independent of the interaction Hamiltonian. Although the gravitational case was discussed By DSW in \cite{Danielson:2022sga}, our result fixes the numerical coefficient of the decoherence rate for the first time.

In the present work, our calculations are limited to the spacetime with a static Killing vector. For future direction on this topic, it will be interesting to extend the local description of decoherence to rotating black holes, where the decoherence from the global perspective was studied in \cite{Gralla:2023oya}. 

\appendix
\section{The Weyl tensor two-point function in de Sitter spacetime}

In this appendix, we present the detailed expression for the Weyl tensor two-point function in de Sitter spacetime obtained in \cite{Kouris:2001hz}. However, as pointed in \cite{Mora:2012kr}, the original expression presented in \cite{Kouris:2001hz} is not the correct one. The error was later fixed by Kouris. One can refer to the corrigendum on the paper by Kouris.

For the de Sitter invariant vacuum, the Weyl tensor two-point function is given by 
\begin{eqnarray}
    W_{abcda'b'c'd'}=\sum_{i=1}^{7} D^{(i)} \Omega^{(i)}_{abcda'b'c'd'}\;,
\end{eqnarray}
where 
\begin{eqnarray}
    D^{(1)}&=&\frac{1}{4\pi^2 L^6} \frac{12}{(z-1)^3}\;,\\
    D^{(2)}&=&\frac{1}{4\pi^2 L^6} \left(\frac{18}{(z-1)^3}-\frac{6}{(z-1)^2}\right)\;,\\
    D^{(3)}&=&\frac{1}{4\pi^2 L^6} \left(-\frac{6}{(z-1)^3}+\frac{6}{(z-1)^2}\right)\;,\\
    D^{(4)}&=&\frac{1}{4\pi^2 L^6} \left(\frac{3}{(z-1)^3}+\frac{3}{(z-1)^2}\right)\;,\\
    D^{(5)}&=&\frac{1}{4\pi^2 L^6} \left(-\frac{3}{2(z-1)^3}+\frac{3}{2(z-1)^2}\right)\;,\\
    D^{(6)}&=&\frac{1}{4\pi^2 L^6} \frac{3}{(z-1)^2}\;,\\
    D^{(7)}&=&\frac{1}{4\pi^2 L^6} \left(\frac{1}{4(z-1)^3}+\frac{3}{4(z-1)^2}\right)\;,
\end{eqnarray}
and 
\begin{eqnarray}
   \Omega^{(i)}_{abcda'b'c'd'}&=&\frac{1}{2}\left(
   S^{(i)}_{[ab][cd][a'b'][c'd']} 
   +S^{(i)}_{[cd][ab][a'b'][c'd']}\right.\nonumber\\
   &&\left.
   +S^{(i)}_{[ab][cd][c'd'][a'b']}  
    +S^{(i)}_{[cd][ab][c'd'][a'b']}
   \right) \;,
\end{eqnarray}
with
\begin{eqnarray}
  S^{(1)}_{abcda'b'c'd'}&=&n_a n_c n_{a'} n_{c'} g_{bd} g_{b'd'} -2 n_a n_c n_{a'} n_{c'} g_{bb'} g_{dd'}\;,\\
  S^{(2)}_{abcda'b'c'd'}&=&\frac{1}{3} n_a n_{c'} \left( g_{bb'} g_{cd'}+g_{bd'}g_{cb'} \right) g_{da'} \;,\\
  S^{(3)}_{abcda'b'c'd'}&=&n_c n_{c'} g_{ab'} g_{bd} g_{a'd'}\;,\\
  S^{(4)}_{abcda'b'c'd'}&=& \left( n_a n_{c} g_{bb'} g_{dd'} g_{a'c'}+ n_{a'} n_{c'} g_{ac} g_{bd'} g_{db'}\right)\nonumber\\&&
  -\frac{1}{2}\left(  n_a n_{c} g_{bd} g_{a'c'} g_{b'd'} + n_{a'} n_{c'} g_{ac} g_{bd} g_{b'd'}\right) \;,\\
  S^{(5)}_{abcda'b'c'd'}&=& \frac{1}{3}\left( g_{ab'} g_{bc'} g_{cd'} g_{da'}+g_{aa'} g_{bb'} g_{cc'} g_{dd'}\right)\;,\\
  S^{(6)}_{abcda'b'c'd'}&=&-g_{ac} g_{bc'} g_{a'd} g_{b'd'} \;,\\
  S^{(7)}_{abcda'b'c'd'}&=&g_{ac} g_{bd} g_{a'c'} g_{b'd'}\;,
\end{eqnarray}
where we have fixed an error in the prefactor of $D^{(i)}$.  

After some tedious calculations, one can finally get the following simple results for the $\Omega^{(i)}$ 
\begin{eqnarray}
    \Omega^{(1)}=-\frac{1}{8}\;,\;
    \Omega^{(2)}=\frac{1}{8}\;,\;
    \Omega^{(3)}=\frac{1}{8}\;,\;
    \Omega^{(4)}=0\;,\;
    \Omega^{(5)}=\frac{1}{4}\;,\;
    \Omega^{(6)}=-\frac{1}{4}\;,\;
    \Omega^{(7)}=\frac{1}{2}\;,\;
\end{eqnarray}
where we have omitted the index of $\Omega^{(i)}_{rtrtr't'r't'}$ for simplicity. With these expressions, one can get the Weyl two-point function given in Eq.\eqref{Weyl_TPF}.

\acknowledgments  The authors would like to thank Hongbao Zhang, Xinghua Wu, Hongji Wei, and Daine L. Danielson for helpful discussions.


\bibliographystyle{JHEP}
\bibliography{biblio.bib}

\providecommand{\href}[2]{#2}\begingroup\raggedright\begin{thebibliography}{10}

\bibitem{Wilson-Gerow:2024ljx}
J.~Wilson-Gerow, A.~Dugad and Y.~Chen, \emph{{Decoherence by warm horizons}}, \href{https://doi.org/10.1103/PhysRevD.110.045002}{\emph{Phys. Rev. D} {\bfseries 110} (2024) 045002} [\href{https://arxiv.org/abs/2405.00804}{{\ttfamily 2405.00804}}].

\bibitem{Penrose:2014nha}
R.~Penrose, \emph{{On the Gravitization of Quantum Mechanics 1: Quantum State Reduction}}, \href{https://doi.org/10.1007/s10701-013-9770-0}{\emph{Found. Phys.} {\bfseries 44} (2014) 557}.

\bibitem{Bassi:2017szd}
A.~Bassi, A.~Gro\ss{}ardt and H.~Ulbricht, \emph{{Gravitational Decoherence}}, \href{https://doi.org/10.1088/1361-6382/aa864f}{\emph{Class. Quant. Grav.} {\bfseries 34} (2017) 193002} [\href{https://arxiv.org/abs/1706.05677}{{\ttfamily 1706.05677}}].

\bibitem{Hu:2008rga}
B.L.~Hu and E.~Verdaguer, \emph{{Stochastic Gravity: Theory and Applications}}, \href{https://doi.org/10.12942/lrr-2008-3}{\emph{Living Rev. Rel.} {\bfseries 11} (2008) 3} [\href{https://arxiv.org/abs/0802.0658}{{\ttfamily 0802.0658}}].

\bibitem{Anastopoulos:2021jdz}
C.~Anastopoulos and B.-L.~Hu, \emph{{Gravitational decoherence: A thematic overview}}, \href{https://doi.org/10.1116/5.0077536}{\emph{AVS Quantum Sci.} {\bfseries 4} (2022) 015602} [\href{https://arxiv.org/abs/2111.02462}{{\ttfamily 2111.02462}}].

\bibitem{Zurek:2003zz}
W.H.~Zurek, \emph{{Decoherence, einselection, and the quantum origins of the classical}}, \href{https://doi.org/10.1103/RevModPhys.75.715}{\emph{Rev. Mod. Phys.} {\bfseries 75} (2003) 715} [\href{https://arxiv.org/abs/quant-ph/0105127}{{\ttfamily quant-ph/0105127}}].

\bibitem{Schlosshauer:2003zy}
M.~Schlosshauer, \emph{{Decoherence, the Measurement Problem, and Interpretations of Quantum Mechanics}}, \href{https://doi.org/10.1103/RevModPhys.76.1267}{\emph{Rev. Mod. Phys.} {\bfseries 76} (2004) 1267} [\href{https://arxiv.org/abs/quant-ph/0312059}{{\ttfamily quant-ph/0312059}}].

\bibitem{Schlosshauer:2019ewh}
M.~Schlosshauer, \emph{{Quantum decoherence}}, \href{https://doi.org/10.1016/j.physrep.2019.10.001}{\emph{Phys. Rept.} {\bfseries 831} (2019) 1} [\href{https://arxiv.org/abs/1911.06282}{{\ttfamily 1911.06282}}].

\bibitem{Danielson:2022tdw}
D.L.~Danielson, G.~Satishchandran and R.M.~Wald, \emph{{Black holes decohere quantum superpositions}}, \href{https://doi.org/10.1142/S0218271822410036}{\emph{Int. J. Mod. Phys. D} {\bfseries 31} (2022) 2241003} [\href{https://arxiv.org/abs/2205.06279}{{\ttfamily 2205.06279}}].

\bibitem{Danielson:2022sga}
D.L.~Danielson, G.~Satishchandran and R.M.~Wald, \emph{{Killing horizons decohere quantum superpositions}}, \href{https://doi.org/10.1103/PhysRevD.108.025007}{\emph{Phys. Rev. D} {\bfseries 108} (2023) 025007} [\href{https://arxiv.org/abs/2301.00026}{{\ttfamily 2301.00026}}].

\bibitem{Belenchia:2018szb}
A.~Belenchia, R.M.~Wald, F.~Giacomini, E.~Castro-Ruiz, v.~Brukner and M.~Aspelmeyer, \emph{{Quantum Superposition of Massive Objects and the Quantization of Gravity}}, \href{https://doi.org/10.1103/PhysRevD.98.126009}{\emph{Phys. Rev. D} {\bfseries 98} (2018) 126009} [\href{https://arxiv.org/abs/1807.07015}{{\ttfamily 1807.07015}}].

\bibitem{Danielson:2021egj}
D.L.~Danielson, G.~Satishchandran and R.M.~Wald, \emph{{Gravitationally mediated entanglement: Newtonian field versus gravitons}}, \href{https://doi.org/10.1103/PhysRevD.105.086001}{\emph{Phys. Rev. D} {\bfseries 105} (2022) 086001} [\href{https://arxiv.org/abs/2112.10798}{{\ttfamily 2112.10798}}].

\bibitem{Gralla:2023oya}
S.E.~Gralla and H.~Wei, \emph{{Decoherence from horizons: General formulation and rotating black holes}}, \href{https://doi.org/10.1103/PhysRevD.109.065031}{\emph{Phys. Rev. D} {\bfseries 109} (2024) 065031} [\href{https://arxiv.org/abs/2311.11461}{{\ttfamily 2311.11461}}].

\bibitem{Li:2024guo}
R.~Li, \emph{{Decoherence of quantum superpositions by Reissner-Nordstr\"om black holes}},  \href{https://arxiv.org/abs/2411.04734}{{\ttfamily 2411.04734}}.

\bibitem{Biggs:2024dgp}
A.~Biggs and J.~Maldacena, \emph{{Comparing the decoherence effects due to black holes versus ordinary matter}},  \href{https://arxiv.org/abs/2405.02227}{{\ttfamily 2405.02227}}.

\bibitem{Danielson:2024yru}
D.L.~Danielson, G.~Satishchandran and R.M.~Wald, \emph{{Local Description of Decoherence of Quantum Superpositions by Black Holes and Other Bodies}},  \href{https://arxiv.org/abs/2407.02567}{{\ttfamily 2407.02567}}.

\bibitem{Unruh:1976db}
W.G.~Unruh, \emph{{Notes on black hole evaporation}}, \href{https://doi.org/10.1103/PhysRevD.14.870}{\emph{Phys. Rev. D} {\bfseries 14} (1976) 870}.

\bibitem{Wald:1995yp}
R.M.~Wald, \emph{{Quantum Field Theory in Curved Space-Time and Black Hole Thermodynamics}}, Chicago Lectures in Physics, University of Chicago Press, Chicago, IL (1995).

\bibitem{Hollands:2014eia}
S.~Hollands and R.M.~Wald, \emph{{Quantum fields in curved spacetime}}, \href{https://doi.org/10.1016/j.physrep.2015.02.001}{\emph{Phys. Rept.} {\bfseries 574} (2015) 1} [\href{https://arxiv.org/abs/1401.2026}{{\ttfamily 1401.2026}}].

\bibitem{Birrell:1982ix}
N.D.~Birrell and P.C.W.~Davies, \emph{{Quantum Fields in Curved Space}}, Cambridge Monographs on Mathematical Physics, Cambridge University Press, Cambridge, UK (1982), \href{https://doi.org/10.1017/CBO9780511622632}{10.1017/CBO9780511622632}.

\bibitem{Polarski:1989bv}
D.~Polarski, \emph{{On the Hawking Effect in De Sitter Space}}, \href{https://doi.org/10.1088/0264-9381/6/5/013}{\emph{Class. Quant. Grav.} {\bfseries 6} (1989) 717}.

\bibitem{Allen:1985wd}
B.~Allen and T.~Jacobson, \emph{{Vector Two Point Functions in Maximally Symmetric Spaces}}, \href{https://doi.org/10.1007/BF01211169}{\emph{Commun. Math. Phys.} {\bfseries 103} (1986) 669}.

\bibitem{Youssef:2010dw}
A.~Youssef, \emph{{Infrared behavior and gauge artifacts in de Sitter spacetime: The photon field}}, \href{https://doi.org/10.1103/PhysRevLett.107.021101}{\emph{Phys. Rev. Lett.} {\bfseries 107} (2011) 021101} [\href{https://arxiv.org/abs/1011.3755}{{\ttfamily 1011.3755}}].

\bibitem{Kouris:2001hz}
S.S.~Kouris, \emph{{The Weyl tensor two point function in de Sitter space-time}}, \href{https://doi.org/10.1088/0264-9381/18/22/316}{\emph{Class. Quant. Grav.} {\bfseries 18} (2001) 4961} [\href{https://arxiv.org/abs/gr-qc/0107064}{{\ttfamily gr-qc/0107064}}].

\bibitem{Mora:2012kr}
P.J.~Mora and R.P.~Woodard, \emph{{Linearized Weyl-Weyl Correlator in a de Sitter Breaking Gauge}}, \href{https://doi.org/10.1103/PhysRevD.85.124048}{\emph{Phys. Rev. D} {\bfseries 85} (2012) 124048} [\href{https://arxiv.org/abs/1202.0999}{{\ttfamily 1202.0999}}].

\bibitem{Allen:1985ux}
B.~Allen, \emph{{Vacuum States in de Sitter Space}}, \href{https://doi.org/10.1103/PhysRevD.32.3136}{\emph{Phys. Rev. D} {\bfseries 32} (1985) 3136}.

\bibitem{Kay:1988mu}
B.S.~Kay and R.M.~Wald, \emph{{Theorems on the Uniqueness and Thermal Properties of Stationary, Nonsingular, Quasifree States on Space-Times with a Bifurcate Killing Horizon}}, \href{https://doi.org/10.1016/0370-1573(91)90015-E}{\emph{Phys. Rept.} {\bfseries 207} (1991) 49}.

\bibitem{Glauber:1963tx}
R.J.~Glauber, \emph{{Coherent and incoherent states of the radiation field}}, \href{https://doi.org/10.1103/PhysRev.131.2766}{\emph{Phys. Rev.} {\bfseries 131} (1963) 2766}.

\bibitem{Zhang:1990fy}
W.-M.~Zhang, D.H.~Feng and R.~Gilmore, \emph{{Coherent States: Theory and Some Applications}}, \href{https://doi.org/10.1103/RevModPhys.62.867}{\emph{Rev. Mod. Phys.} {\bfseries 62} (1990) 867}.

\bibitem{2012JPhA...45x4002S}
B.C.~{Sanders}, \emph{{Review of entangled coherent states}}, \href{https://doi.org/10.1088/1751-8113/45/24/244002}{\emph{Journal of Physics A Mathematical General} {\bfseries 45} (2012) 244002} [\href{https://arxiv.org/abs/1112.1778}{{\ttfamily 1112.1778}}].

\bibitem{Gibbons:1977mu}
G.W.~Gibbons and S.W.~Hawking, \emph{{Cosmological Event Horizons, Thermodynamics, and Particle Creation}}, \href{https://doi.org/10.1103/PhysRevD.15.2738}{\emph{Phys. Rev. D} {\bfseries 15} (1977) 2738}.

\bibitem{Spradlin:2001pw}
M.~Spradlin, A.~Strominger and A.~Volovich, \emph{{Les Houches lectures on de Sitter space}},  in \emph{{Les Houches Summer School: Session 76: Euro Summer School on Unity of Fundamental Physics: Gravity, Gauge Theory and Strings}}, pp.~423--453, 10, 2001 [\href{https://arxiv.org/abs/hep-th/0110007}{{\ttfamily hep-th/0110007}}].

\bibitem{Saharian:2013yya}
A.A.~Saharian, A.S.~Kotanjyan and H.A.~Nersisyan, \emph{{Electromagnetic two-point functions and Casimir densities for a conducting plate in de Sitter spacetime}}, \href{https://doi.org/10.1016/j.physletb.2013.11.041}{\emph{Phys. Lett. B} {\bfseries 728} (2014) 141} [\href{https://arxiv.org/abs/1307.5536}{{\ttfamily 1307.5536}}].

\bibitem{Higuchi:2001uv}
A.~Higuchi and S.S.~Kouris, \emph{{The Covariant graviton propagator in de Sitter space-time}}, \href{https://doi.org/10.1088/0264-9381/18/20/311}{\emph{Class. Quant. Grav.} {\bfseries 18} (2001) 4317} [\href{https://arxiv.org/abs/gr-qc/0107036}{{\ttfamily gr-qc/0107036}}].

\end{thebibliography}\endgroup
\end{document}